\documentclass[english,notitlepage,nofootinbib]{revtex4-1}
\usepackage[T1]{fontenc}
\usepackage[latin9]{inputenc}
\usepackage{graphicx}

\makeatletter

\providecommand{\tabularnewline}{\\}

\makeatother

\usepackage{babel}
\begin{document}

\title{Minima of the Scalar Potential in the Type II Seesaw Model: \\
From Local to Global }

\date{\today}

\author{Xun-Jie Xu, }

\affiliation{Max-Planck-Institut f\"ur Kernphysik, Postfach 103980, D-69029 Heidelberg,
Germany.}
\begin{abstract}
The Type II seesaw model requires that its scalar doublet $H$ and
triplet $\Delta$ get specific patterns of VEVs $\langle H\rangle\propto(0,1)^{T}$
and $\langle\Delta\rangle\propto(0,0,1)^{T}$ to accommodate neutrino
masses. However, other types of minima could also exist in the scalar
potential, which may strongly contradict to experimental observations.
This paper studies when the minimum at $\langle H\rangle\propto(0,1)^{T}$
and $\langle\Delta\rangle\propto(0,0,1)^{T}$ will be global and
finds the necessary and sufficient condition for that, assuming that
the lepton number violating term $H^{2}\Delta$ in the potential is
perturbatively small.
\end{abstract}
\maketitle

\section{Introduction}

As a well-tested  theory of particle physics, the Standard Model
(SM) contains only one scalar boson, i.e. the Higgs boson, which has
been found at the Large Hadron Collider (LHC) since 2012 \cite{Aad:2012tfa,Chatrchyan:2012xdj}.
To go beyond the SM, it is possible that the SM might be extended
by new scalar fields such as an $SU(2)_{L}$ singlet, or another $SU(2)_{L}$
doublet (known as the two-Higgs-doublet model \cite{Branco:2011iw}),
or an $SU(2)_{L}$ triplet, etc. Actually for the triplet extension,
 neutrino oscillation which has been well established \cite{Capozzi:2013csa,Forero:2014bxa,Gonzalez-Garcia:2014bfa}
by many experiments and indicates that neutrinos have tiny masses
could be an important hint. Adding a triplet with hypercharge $Y=1$
to the SM can naturally generate\footnote{Another case $Y=0$ (a real triplet) has also been studied in the
literature, see for e.g. \cite{Gunion:1989ci,Blank:1997qa,Forshaw:2001xq,Forshaw:2003kh,Chen:2006pb,Chankowski:2006hs,SekharChivukula:2007gi,Chen:2008jg,FileviezPerez:2008bj}.
But the real triplet model can not generate neutrino masses. We will
not discuss it in this paper.} small neutrino masses via the Type II seesaw mechanism \cite{Konetschny:1977bn,Cheng:1980qt,Schechter:1980gr}
so this extension is often referred to as the Type II seesaw model,
which has been widely studied in many references \cite{Bonilla:2015eha,Han:2015hba,Han:2015sca,Das:2016bir,Haba:2016zbu,Chabab:2015nel,Bahrami:2015mwa,Chabab:2014ara,Chen:2014lla,Chen:2013dh,Dev:2013ff,Arbabifar:2012bd,Aoki:2012jj,Chao:2012mx,Chun:2012jw,Akeroyd:2012ms,Aoki:2012yt,Akeroyd:2012nd,Arhrib:2011vc,Aoki:2011pz,Akeroyd:2011ir,Akeroyd:2011zza}
recently.

In the Type II seesaw model, both the doublet $H$ and the triplet
$\Delta$ acquire nonzero vacuum expectation values (VEV),
\begin{equation}
\langle H\rangle\propto(0,1)^{T},\thinspace\langle\Delta\rangle\propto(0,0,1)^{T}.\label{eq:t2-63}
\end{equation}
Note that other possibilities such as $\langle\Delta\rangle\propto(1,0,0)^{T}$,
$(0,1,0)^{T}$ (or more generally $(x,y,z)^{T}$ with arbitrary nonzero
numbers $x,$ $y$ and $z$) would break the electromagnetic symmetry
$U(1)_{{\rm em}}$. Besides, neutrinos obtain Majorana masses only
when $\langle\Delta\rangle\propto(0,0,1)^{T}$. So in accord with
experimental facts, the VEVs of $H$ and $\Delta$ have to be in the
form of Eq.\,(\ref{eq:t2-63}). Usually such VEVs are assumed to
be obtained in the scalar potential of the type II seesaw model.
Indeed, if one tries to solve the minimization equation of the potential
$V(H,\Delta)$ 
\begin{equation}
\frac{\partial V}{\partial H}=0,\thinspace\frac{\partial V}{\partial\Delta}=0,\label{eq:t2-64}
\end{equation}
then Eq.\,(\ref{eq:t2-63}) is one of its solutions. So conventionally
they are directly used in most studies on this model.

However, other solutions may also exist, corresponding to some different
minima. In other words, despite that the VEVs in Eq.\,(\ref{eq:t2-63})
are required by experimental facts, the most general scalar potential
does not necessarily lead to this result. Actually, in this paper
we find that other minima which correspond to VEVs very different
from Eq.\,(\ref{eq:t2-63}) do exist and sometimes they can be the
global minimum, i.e. the deepest point of the potential. 

If there were some minima deeper than the one at Eq.\,(\ref{eq:t2-63}),
then the vacuum corresponding to Eq.\,(\ref{eq:t2-63}) could be
unstable and would decay into a deeper minimum. Consequently, the
Type II seesaw model would fail to describe the real world where we
have unbroken $U(1)_{{\rm em}}$ and very light neutrinos.

Therefore we would like to know when the minimum at Eq.\,(\ref{eq:t2-63})
becomes the global minimum. In this paper we will analyze the potential
and find all the solutions of the equation of minimization analytically.
It turns out that there are only four different types of minima. Only
one of them respects $U(1)_{{\rm em}}$ and generates tiny neutrino
masses but it is not always the deepest minimum. By comparing the
four minima, we obtain the condition of this minimum being the deepest
one. If the parameters of the potential are constrained by this condition,
the globalness of this minimum can be guaranteed and the vacuum usually
considered in the Type II seesaw model will be stable.

The paper is organized as follows. In Sec.\,\ref{sec:The-Type-II}
we will give a brief introduction of the Type II Seesaw model and
discuss on some issues related to the potential such as the smallness
of $\mu$ (the coefficient of the lepton-number-violating term), physical
equivalence of some minima, complex phases in the scalar fields, etc.
Then the analysis is based on the expansion in small $\mu$. We first
try to solve the minimization equation at the Leading Order (LO) with
$\mu=0$ in Sec.\,\ref{sec:Local-Minima-At}, where all the solutions
will be analytically found and numerically verified. Then in Sec.\,\ref{sec:From-local-to}
we compare these solutions at the LO to find out when the desired
vacuum  is the deepest minimum. This will produce some constraints
on the parameters of the potential, which is verified in Fig.\,\ref{fig:togloabl}.
Following the LO calculation, we go further to the Next-to-Leading
Order, presented in Sec.\,\ref{sec:NLO}. Predictions of the scalar
mass spectrum from the new constraints are studied in Sec.\,\ref{sec:Prediction}.
Finally, we conclude in Sec.\,\ref{sec:Discussion-and-Conclusion}.

\section{\label{sec:The-Type-II}The Type II Seesaw Model}

In the Type II seesaw model, there are two scalar fields, an $SU(2)_{L}$
doublet (i.e. the SM Higgs) and an $SU(2)_{L}$ triplet $\Delta$
with hypercharges $Y_{H}=\frac{1}{2}$ and $Y_{\Delta}=1$ respectively,
\begin{equation}
H=(\phi^{+},\phi^{0})^{T},\thinspace\Delta=(\delta^{++},\delta^{+},\delta^{0})^{T}.\label{eq:t2-4}
\end{equation}
Under an arbitrary $SU(2)_{L}$ transformation $\exp(i\epsilon_{a}t_{a})\in SU(2)_{L}$
where $t_{a}$ ($a=1,2,3$) are the three generators of $SU(2)_{L}$,
the doublet and triplet are transformed as 
\begin{equation}
H\rightarrow\exp(i\epsilon_{a}T_{a}^{j=1/2})H,\thinspace\Delta\rightarrow\exp(i\epsilon_{a}T_{a}^{j=1})\Delta.\label{eq:t2-5}
\end{equation}
Here $T_{a}^{j=1/2}$ and $T_{a}^{j=1}$ are matrix representations
of $t_{a}$ for the $SU(2)_{L}$ spin $j=1/2$ and $j=1$, respectively.\footnote{More explicitly, for $a=3$ we have $T_{3}^{j=1/2}={\rm diag}(1/2,-1/2)$
and $T_{3}^{j=1}={\rm diag}(1,0,-1)$. For $a=1,\thinspace2$, $T_{a}^{j=1/2}$and
$T_{a}^{j=1}$ can be computed via the well-known raising and lowering
operators of the $SU(2)$ algebra. The result is $T_{a}^{j=1/2}=\sigma_{a}/2$
where $\sigma_{a}$'s are Pauli matrices, $T_{1}^{j=1}=\frac{1}{\sqrt{2}}\left(\begin{array}{ccc}
0 & 1 & 0\\
1 & 0 & 1\\
0 & 1 & 0
\end{array}\right)$, $T_{2}^{j=1}=\frac{1}{\sqrt{2}}\left(\begin{array}{ccc}
0 & -i & 0\\
i & 0 & -i\\
0 & i & 0
\end{array}\right)$.} However the above representation for $\Delta$ is not convenient
in constructing $SU(2)_{L}$ invariants, as the CG coefficients are
involved. A more convenient way (and also more conventional in the
literature) is to use the matrix form

\begin{equation}
\Delta=\left(\begin{array}{cc}
\delta^{+}/\sqrt{2} & \delta^{++}\\
\delta^{0} & -\delta^{+}/\sqrt{2}
\end{array}\right),\label{eq:t2-7}
\end{equation}
which transforms as a traceless rank-2 tensor of $SU(2)_{L}$, 
\begin{equation}
\Delta\rightarrow\exp(i\epsilon_{a}T_{a}^{j=1/2})\Delta\exp(-i\epsilon_{a}T_{a}^{j=1/2}).\label{eq:t2-6}
\end{equation}

The scalar potential of the model in terms of $H$ and $\Delta$ {[}in
its matrix form given by Eq.\,(\ref{eq:t2-7}){]} can be written
as

\begin{equation}
V=-m^{2}H^{\dagger}H+M^{2}{\rm Tr}\left[\Delta^{\dagger}\Delta\right]+V_{H}+V_{\Delta}+V_{H\Delta},\label{eq:t2}
\end{equation}
where
\begin{eqnarray}
 & V_{H}= & \frac{\lambda}{4}(H^{\dagger}H)^{2},\label{eq:t2-1}\\
 & V_{\Delta}= & \lambda_{2}\left[{\rm Tr}\Delta^{\dagger}\Delta\right]^{2}+\lambda_{3}{\rm Tr}\left[\Delta^{\dagger}\Delta\right]^{2},\label{eq:t2-2}\\
 & V_{H\Delta}= & \left[\mu H^{T}i\sigma_{2}\Delta^{\dagger}H+h.c.\right]+\lambda_{1}(H^{\dagger}H){\rm Tr}\left[\Delta^{\dagger}\Delta\right]+\lambda_{4}H^{\dagger}\Delta\Delta^{\dagger}H.\label{eq:t2-3}
\end{eqnarray}
The parameter $\mu$ in the trilinear term ($H$-$\Delta$-$H$) 
could be complex. However, since its complex phase can always be absorbed
into $\Delta$ and $H$, we can set it real. Therefore all parameters
($m$, $M$, $\mu$, $\lambda$, $\lambda_{1,2,3,4}$) are real in
the potential.

The Type II seesaw model generates neutrino masses via the Yukawa
interaction of $\Delta$ with the left-handed lepton doublet $L$,
\begin{equation}
{\cal L}_{{\rm Yukawa}}\supset Y_{\nu}L^{T}i\sigma_{2}\Delta L+h.c.\thinspace,\label{eq:t2-59}
\end{equation}
where $Y_{\nu}$ is the corresponding Yukawa coupling (matrix). The
scalar potential is expected to lead to the following VEVs,  
\begin{equation}
\langle H\rangle=\frac{v_{H}}{\sqrt{2}}\left(\begin{array}{c}
0\\
1
\end{array}\right),\thinspace\langle\Delta\rangle=\frac{v_{\Delta}}{\sqrt{2}}\left(\begin{array}{cc}
0 & 0\\
1 & 0
\end{array}\right).\label{eq:t2-60}
\end{equation}
Therefore after symmetry breaking, neutrinos obtain masses approximately
(assuming $M^{2}\gg m^{2}$) of order 
\begin{equation}
m_{\nu}\sim Y_{\nu}\mu v_{H}^{2}/M^{2}.\label{eq:t2-61}
\end{equation}
Taking $m_{\nu}\sim0.05{\rm eV}$ and $v_{H}\approx246{\rm GeV}$,
the above relation can be written as 
\begin{equation}
\mu\sim0.8{\rm eV}\frac{1}{Y_{\nu}}\left(\frac{M}{1{\rm TeV}}\right)^{2},\label{eq:t2-62}
\end{equation}
which implies that if $M$ is at the TeV scale and $Y_{\nu}\sim O(1)$
then $\mu$ is of order $O({\rm eV})$, much less than $m$ or $M$.
And even if $M$ is hundreds of TeV, $\mu\ll m,\thinspace M$ still
holds. Note that if $\mu=0$, i.e. we turn off the trilinear term,
the lepton number is conserved. So $\mu$ could be naturally very
small as a result of symmetry. In this paper, we will assume that
$\mu$ is perturbatively small, which means it is small enough so
that all calculations based on the expansion in $\mu$ are valid.
This can be guaranteed if $\mu\ll m,\thinspace M$ and there is no
fine-tuning in the quartic couplings $\lambda$ and $\lambda_{1,2,3,4}$.
In this paper, the potential will be analyzed first at the  LO assuming
$\mu=0$ and then at the  NLO which holds for nonzero $\mu$ up to
$O(\mu^{2})$. 

Before getting to the detailed analysis, we would like to introduce
some techniques that will simplify the minimization of the scalar
potential, independent of whether $\mu$ is small or not. Note that
even in the SM, the minimum of the Higgs potential is not unique due
to the $SU(2)_{L}\times U(1)_{Y}$ invariance. For example, the value
of the Higgs potential at $H=(0,v/\sqrt{2})^{T}$ and the value at
$H=\exp(i\epsilon_{a}T_{a}^{j=1/2})(0,v/\sqrt{2})^{T}$ for arbitrary
$\epsilon_{a}$'s are always equal, implying that there are infinite
numbers of minima. However, these minima are physically equivalent
since they are connected via gauge transformations. In a similar way,
the Type II seesaw model also has infinite numbers of physically equivalent
minima due to the $SU(2)_{L}\times U(1)_{Y}$  invariance. Actually,
the degrees of freedom of the  transformations among these equivalent
minima correspond to Goldstone bosons. In the SM, the unitarity gauge
is sometimes used to avoid the explicit appearance of Goldstone bosons,
i.e. to absorb them into the gauge fields. In the unitarity gauge,
the Higgs potential of the SM becomes particularly simple. In the
Type II seesaw model we deal with it in the same way. We can always
transform $H$ and $\Delta$ to the following form
\begin{equation}
H=\left(\begin{array}{c}
0\\
h/\sqrt{2}
\end{array}\right),\thinspace\Delta=\left(\begin{array}{cc}
ye^{i\phi_{y}}/\sqrt{2} & ze^{i\phi_{z}}\\
xe^{i\phi_{x}} & -ye^{i\phi_{y}}/\sqrt{2}
\end{array}\right),\label{eq:t2-7-1}
\end{equation}
by $H\rightarrow\exp(i\epsilon_{a}T_{a}^{j=1/2})H$ and $\Delta\rightarrow\exp(i\epsilon_{a}T_{a}^{j=1/2})\Delta\exp(-i\epsilon_{a}T_{a}^{j=1/2})$.
All the fields ($h$, $x$, $y$, $z$, $\phi_{x}$, $\phi_{y}$,
$\phi_{z}$) in Eq.(\ref{eq:t2-7-1}) are real fields. So the potential
becomes 
\begin{equation}
V=-\frac{m^{2}}{2}h^{2}+M^{2}(x^{2}+y^{2}+z^{2})+V_{H}+V_{\Delta}+V_{H\Delta},\label{eq:t2-8}
\end{equation}
where
\begin{eqnarray}
 & V_{H}= & \frac{\lambda}{16}h^{4},\label{eq:t2-1-1}\\
 & V_{\Delta}= & \lambda_{2}(x^{2}+y^{2}+z^{2})^{2}+\lambda_{3}\left[x^{4}+z^{4}+y^{2}\left(2x^{2}+\frac{y^{2}}{2}+2z^{2}-2xz\cos(\phi_{x}-2\phi_{y}+\phi_{z})\right)\right],\label{eq:t2-2-1}\\
 & V_{H\Delta}= & -\mu xh^{2}\cos\phi_{x}+\frac{\lambda_{1}}{2}h^{2}(x^{2}+y^{2}+z^{2})+\frac{\lambda_{4}}{4}h^{2}(2x^{2}+y^{2}).\label{eq:t2-3-1}
\end{eqnarray}

From the above expressions we can see that actually the complex phases
$\phi_{x,y,z}$ in $\Delta$ are not important when the minimization
is concerned, explained as follows. The complex phases only appear
in two terms, $xh^{2}\cos\phi_{x}$ in $V_{H\Delta}$ and $y^{2}xz\cos(\phi_{x}-2\phi_{y}+\phi_{z})$
in $V_{\Delta}$. For simplicity, consider a function $f(A,B,\alpha,\beta)=g(A,B)+A\cos\alpha+B\cos\beta$.
 If a minimum (no matter local or global) of $f$ appears somewhere
with $A\neq0$ then $\cos\alpha$ at this minimum should be $1$ or
$-1$. If it appears with $A=0$, then the value of $\cos\alpha$
is irrelevant. That is to say, there can not be a minimum with $|\cos\alpha|\neq1$
unless the value of $f$ at the minimum does not depend on $\alpha$.
Similar arguments also apply to $\beta$. So we can always set $|\cos\alpha|,\thinspace|\cos\beta|=1$
in order to find the minimum. Taking $\alpha=\phi_{x}$ and $\beta=\phi_{x}-2\phi_{y}+\phi_{z}$
we can get the conclusion that a minimum of (\ref{eq:t2-8}) always
has $|\cos\phi_{x}|=|\cos(\phi_{x}-2\phi_{y}+\phi_{z})|=1$ unless
the value of $V$ at the minimum is independent of these phases. There
are 4 cases $(\cos\alpha,\cos\beta)=$$(1,1)$, $(1,-1)$, $(-1,1)$
and $(-1,-1)$, which can be obtained by setting $(\phi_{x},\phi_{y},\phi_{z})=(0,0,0)$,
$(0,0,\pi)$, $(\pi,0,\pi)$ and $(\pi,0,0)$, respectively. However,
from (\ref{eq:t2-7-1}) we can see that $\Delta$ is invariant under
the transformation $(x,\phi_{x})\rightarrow(-x,\phi_{x}-\pi)$ or
$(z,\phi_{z})\rightarrow(-z,\phi_{z}-\pi)$. Therefore we can set
$(\phi_{x},\phi_{y},\phi_{z})=(0,0,0)$ and take $x,\thinspace y,\thinspace z\in(-\infty,\thinspace+\infty)$
to find any local or global  minimum.

\section{\label{sec:Local-Minima-At}Local Minima At The Leading Order}

In this section we analyze the potential at the LO (i.e. assuming
$\mu=0$) to find all the local minima. We find that there are only
four types of minima  at the LO, which are listed in Tab.\ref{tab:All-local-minima}.
In the main part of this section, we will show how to analytically
solve the equation of minimization to find them, and then numerically
verify both the correctness and exhaustiveness of these analytic solutions.

\subsection{Analytical solutions}

\begin{table}
\centering

\begin{tabular}{|c|c|c|c|c|}
\hline 
Solutions & Type A & Type B & Type C & Type D\tabularnewline
\hline 
\rule[-5ex]{0pt}{10ex}  $h^{2}$ & $\frac{4m^{2}}{\lambda}$ & $-\frac{2\left(M^{2}\lambda_{1}+2m^{2}\left(\lambda_{2}+\lambda_{3}\right)\right)}{\lambda_{1}^{2}-\lambda\left(\lambda_{2}+\lambda_{3}\right)}$ & $-\frac{2\left(2m^{2}\left(\lambda_{2}+\lambda_{3}\right)+M^{2}\left(\lambda_{1}+\lambda_{4}\right)\right)}{-\lambda\left(\lambda_{2}+\lambda_{3}\right)+\left(\lambda_{1}+\lambda_{4}\right){}^{2}}$ & $\frac{2\lambda_{3}\left(2m^{2}\left(2\lambda_{2}+\lambda_{3}\right)+M^{2}\left(2\lambda_{1}+\lambda_{4}\right)\right)}{\lambda_{3}\left(-2\lambda_{1}^{2}+\lambda\left(2\lambda_{2}+\lambda_{3}\right)\right)-2\lambda_{1}\lambda_{3}\lambda_{4}-\left(\lambda_{2}+\lambda_{3}\right)\lambda_{4}^{2}}$\tabularnewline
\rule[-5ex]{0pt}{10ex}  $x^{2}$ & 0 & 0 & $\frac{M^{2}\lambda+2m^{2}\left(\lambda_{1}+\lambda_{4}\right)}{2\left(-\lambda\left(\lambda_{2}+\lambda_{3}\right)+\left(\lambda_{1}+\lambda_{4}\right){}^{2}\right)}$ & $\frac{\left(M^{2}\lambda_{1}+2m^{2}\lambda_{2}\right)\lambda_{4}+\lambda_{3}\left(M^{2}\lambda+2m^{2}\left(\lambda_{1}+\lambda_{4}\right)\right)}{4\lambda_{1}^{2}\lambda_{3}-2\lambda\lambda_{3}\left(2\lambda_{2}+\lambda_{3}\right)+4\lambda_{1}\lambda_{3}\lambda_{4}+2\left(\lambda_{2}+\lambda_{3}\right)\lambda_{4}^{2}}$\tabularnewline
$y^{2}$ & 0 & 0 & 0 & 0\tabularnewline
\rule[-5ex]{0pt}{10ex}  $z^{2}$ & 0 & $\frac{M^{2}\lambda+2m^{2}\lambda_{1}}{2\left(\lambda_{1}^{2}-\lambda\left(\lambda_{2}+\lambda_{3}\right)\right)}$ & 0 & $\frac{\left(M^{2}\lambda+2m^{2}\lambda_{1}\right)\lambda_{3}-\lambda_{4}\left(2m^{2}\lambda_{2}+M^{2}\left(\lambda_{1}+\lambda_{4}\right)\right)}{4\lambda_{1}^{2}\lambda_{3}-2\lambda\lambda_{3}\left(2\lambda_{2}+\lambda_{3}\right)+4\lambda_{1}\lambda_{3}\lambda_{4}+2\left(\lambda_{2}+\lambda_{3}\right)\lambda_{4}^{2}}$\tabularnewline
\hline 
\rule[-5ex]{0pt}{10ex}  $V_{{\rm min}}$ & $-\frac{m^{4}}{\lambda}$ & $-\frac{M^{4}\lambda+4m^{2}M^{2}\lambda_{1}+4m^{4}\left(\lambda_{2}+\lambda_{3}\right)}{4\left(-\lambda_{1}^{2}+\lambda\left(\lambda_{2}+\lambda_{3}\right)\right)}$ & $-\frac{M^{4}\lambda+4m^{2}\left(m^{2}\left(\lambda_{2}+\lambda_{3}\right)+M^{2}\left(\lambda_{1}+\lambda_{4}\right)\right)}{4\left(\left(\lambda_{1}+\lambda_{4}\right){}^{2}-\lambda\left(\lambda_{2}+\lambda_{3}\right)\right)}$ & $\frac{4m^{4}\lambda_{3}^{2}-M^{4}\lambda_{4}^{2}+2\lambda_{3}\left(M^{4}\lambda+2m^{2}\left(2m^{2}\lambda_{2}+M^{2}\left(2\lambda_{1}+\lambda_{4}\right)\right)\right)}{8\lambda_{1}^{2}\lambda_{3}-4\lambda\lambda_{3}\left(2\lambda_{2}+\lambda_{3}\right)+8\lambda_{1}\lambda_{3}\lambda_{4}+4\left(\lambda_{2}+\lambda_{3}\right)\lambda_{4}^{2}}$\tabularnewline
\hline 
\end{tabular}

\caption{\label{tab:All-local-minima}All local minima of the scalar potential
at LO, computed by solving Eq.\,(\ref{eq:t2-13}).}
\end{table}

As previously mentioned, we can set $(\phi_{x},\phi_{y},\phi_{z})=(0,0,0)$
without loss of generality in finding all the minima of the potential.
So we consider the potential $V_{0}$ (the subscript $0$ is to remind
us that it is at the LO) with $(\phi_{x},\phi_{y},\phi_{z})=(0,0,0)$
and $\mu=0$, which is 
\begin{eqnarray}
V_{0} & = & -\frac{m^{2}}{2}h^{2}+M^{2}(x^{2}+y^{2}+z^{2})+\frac{\lambda}{16}h^{4}\nonumber \\
 &  & +\lambda_{2}(x^{2}+y^{2}+z^{2})^{2}+\lambda_{3}\left[x^{4}+z^{4}+y^{2}\left(2x^{2}+\frac{y^{2}}{2}+2z^{2}-2xz\right)\right]\nonumber \\
 &  & +\frac{\lambda_{1}}{2}h^{2}(x^{2}+y^{2}+z^{2})+\frac{\lambda_{4}}{4}h^{2}(2x^{2}+y^{2}).\label{eq:t2-9}
\end{eqnarray}

Next we are going to find solutions of the equation

\begin{equation}
\frac{\partial V_{0}}{\partial h}=\frac{\partial V_{0}}{\partial x}=\frac{\partial V_{0}}{\partial y}=\frac{\partial V_{0}}{\partial z}=0.\label{eq:t2-13}
\end{equation}

Note that $V_{0}$ is an even function of $y$ and $h$, i.e. $V_{0}=V_{0}(h^{2},y^{2})$,
since $y$ and $h$ only appear in the form $y^{2}$ and $h^{2}$
in Eq.\,(\ref{eq:t2-9}). An important implication is that $y=0$
is a solution of $\partial V_{0}/\partial y=0$. Similar conclusion
also holds for $h=0$. This can be seen by 
\begin{equation}
\frac{\partial V_{0}}{\partial y}=\frac{\partial V_{0}(y^{2})}{\partial y^{2}}\frac{dy^{2}}{dy}=2y\frac{\partial V_{0}}{\partial y^{2}},\label{eq:t2-10}
\end{equation}
which means the derivative would be zero if $y$ is zero. If $y=0$,
then $V_{0}$ further becomes an even function of $x$ and $z$. In
this case $x=0$ and $z=0$ are solutions of the corresponding minimization
equations. These zero solutions imply that zeros in $(h,x,y,z)$ are
very common for the local minima of $V_{0}$.

We first focus on the case $y=0$. In this case we only need to solve
$\partial_{h}V_{0}=\partial_{x}V_{0}=\partial_{z}V_{0}=0$ since $\partial_{y}V_{0}$
is already zero. Although $h=0$ is a solution of $\partial_{h}V_{0}=0$
it can be proven that $h=0$ leads to a saddle point, not a local
minimum. The proof is given as follows.

For $h=y=0$, the potential is simplified to 
\begin{equation}
V_{xz}=M^{2}(x^{2}+z^{2})+\lambda_{x}x^{4}+\lambda_{z}z^{4}+2\lambda_{xz}x^{2}z^{2},\label{eq:t2-11}
\end{equation}
where $\lambda_{x}$, $\lambda_{z}$ and $\lambda_{xz}$ are some
linear combinations of $\lambda_{2}$ and $\lambda_{3}$. We first
prove that this potential has only one minimum $x=z=0$. If $\lambda_{xz}\geq0$,
this is obvious. If $\lambda_{xz}<0$, then from the bounded-from-below
(BFB) condition we have $\lambda_{x}\lambda_{z}-\lambda_{xz}^{2}\geq0$,
to be used later. There could be four types of minima, (i) $x\neq0$,
$z=0$; (ii) $x=0$, $z\neq0$; (iii) $x\neq0$, $z\neq0$; (iv) $x=0$,
$z=0$, discussed below.
\begin{description}
\item [{(i)}] This can be excluded immediately since $V_{xz}(z=0)=M^{2}x^{2}+\lambda_{x}x^{4}$
has only one minimum $x=0$.
\item [{(ii)}] Similar to (i), this can also be excluded.
\item [{(iii)}] By solving the equation of minimization, the minimum can
be computed $(x^{2},z^{2})=\frac{M^{2}}{\lambda_{x}\lambda_{z}-\lambda_{xz}^{2}}(\lambda_{xz}-\lambda_{z},\lambda_{xz}-\lambda_{x})$.
Note that $\lambda_{x}\lambda_{z}-\lambda_{xz}^{2}>0$ and $\lambda_{xz}<0$,
which implies that it is a negative solution, contradict to the fact
that $x$ and $z$ here are squared.
\item [{(iv)}] This remains as the only possible minimum of $V_{xz}$.
\end{description}
Thus we have proved that the only minimum of $V_{xz}$ is $x=z=0$.
So $(h,x,y,z)=(0,0,0,0)$ is a solution of Eq.\,(\ref{eq:t2-13})
with $h=y=0$, and thus the only would-be minimum of $V_{0}$ for
$h=0$. However, this point is actually a saddle point, because
we can find a direction in which the potential goes down. When $h$
is increased a little from the point while $x$, $y$, $z$ are kept
zero, then potential (\ref{eq:t2-9}) would be $-\frac{m^{2}}{2}h^{2}+\frac{\lambda}{16}h^{4}$
which goes down as long as $h^{2}$ does not excess $\frac{4m^{2}}{\lambda}$.
So the only would-be minimum for $h=0$ turns out to be a saddle point. 

Therefore we only need to consider $h\neq0$. Again, there are four
possible solutions of Eq.\,(\ref{eq:t2-13}), depending on whether
$x$ and $z$ equal to zero, discussed  below.  

{\bf Case A: }$x=z=0$

Only $h$ is nonzero in this case. Note that $\partial_{x}V_{0},\thinspace\partial_{y}V_{0}$
and $\partial_{z}V_{0}$ are zero at $x=y=z=0$. Only $\partial_{h}V_{0}=0$
remains to be solved. Since $h$ is nonzero, we will solve $\partial V_{0}/\partial h^{2}=0$
instead. The potential at $x=y=z=0$ is 
\begin{equation}
V_{0}|_{x=y=z=0}=-\frac{m^{2}}{2}h^{2}+\frac{\lambda}{16}h^{4}.\label{eq:t2-12}
\end{equation}
So $\partial V_{0}/\partial h^{2}=0$ gives $h^{2}=\frac{4m^{2}}{\lambda}$.
Therefore, we find a solution of Eq.\,(\ref{eq:t2-13}), named as
the Type A solution. Correspondingly, the minimum will be referred
to as the Type A minimum and the vacuum at this minimum is called
the Type A vacuum. The potential value at this minimum is 
\begin{equation}
V_{{\rm min}=}^{A}-\frac{m^{4}}{\lambda}.\label{eq:0626}
\end{equation}

{\bf Case B: }$x=0$, $z\neq0$

Only $h$ and $z$ are nonzero in this case. The potential at $x=y=0$
is 
\begin{eqnarray}
V_{0} & |_{x=y=0}= & -\frac{m^{2}}{2}h^{2}+M^{2}z^{2}+\frac{\lambda}{16}h^{4}+(\lambda_{2}+\lambda_{3})z^{4}+\frac{\lambda_{1}}{2}h^{2}z^{2}.\label{eq:t2-14}
\end{eqnarray}
From the equation $\partial V_{0}/\partial h^{2}=0$ and $\partial V_{0}/\partial z^{2}=0$
we can get the solution 
\begin{equation}
h^{2}=-\frac{2\left(M^{2}\lambda_{1}+2m^{2}\left(\lambda_{2}+\lambda_{3}\right)\right)}{\lambda_{1}^{2}-\lambda\left(\lambda_{2}+\lambda_{3}\right)},\label{eq:t2-15}
\end{equation}
\begin{equation}
z^{2}=\frac{M^{2}\lambda+2m^{2}\lambda_{1}}{2\left(\lambda_{1}^{2}-\lambda\left(\lambda_{2}+\lambda_{3}\right)\right)}.\label{eq:t2-16}
\end{equation}
We call it the Type B solution. The potential value at this minimum
is 
\begin{equation}
V_{{\rm min}=}^{B}-\frac{M^{4}\lambda+4m^{2}M^{2}\lambda_{1}+4m^{4}\left(\lambda_{2}+\lambda_{3}\right)}{4\left(\lambda\left(\lambda_{2}+\lambda_{3}\right)-\lambda_{1}^{2}\right)}.\label{eq:0626-1}
\end{equation}

{\bf Case C: }$x\neq0$, $z=0$

Only $h$ and $x$ are nonzero in this case. The potential at $z=y=0$
is 
\begin{eqnarray}
V_{0} & |_{z=y=0}= & -\frac{m^{2}}{2}h^{2}+M^{2}x^{2}+\frac{\lambda}{16}h^{4}+(\lambda_{2}+\lambda_{3})z^{4}+\frac{\lambda_{1}+\lambda_{4}}{2}h^{2}x^{2}.\label{eq:t2-14-1}
\end{eqnarray}
Note that the above potential is very similar to Case B. If we replace
$x$ with $z$ and $\lambda_{1}$ with $\lambda_{1}+\lambda_{4}$,
then it goes back to Eq. (\ref{eq:t2-14}). So we do not need to solve
the equation again, the solution can be obtained from Case B with
the simple replacement

\begin{equation}
h^{2}=-\frac{2\left(2m^{2}\left(\lambda_{2}+\lambda_{3}\right)+M^{2}\left(\lambda_{1}+\lambda_{4}\right)\right)}{-\lambda\left(\lambda_{2}+\lambda_{3}\right)+\left(\lambda_{1}+\lambda_{4}\right){}^{2}},\label{eq:t2-15-1}
\end{equation}
\begin{equation}
x^{2}=\frac{M^{2}\lambda+2m^{2}\left(\lambda_{1}+\lambda_{4}\right)}{2\left(-\lambda\left(\lambda_{2}+\lambda_{3}\right)+\left(\lambda_{1}+\lambda_{4}\right){}^{2}\right)}.\label{eq:t2-16-1}
\end{equation}
We call it the Type C solution. The potential value at this minimum
is 
\begin{equation}
V_{{\rm min}=}^{C}-\frac{M^{4}\lambda+4m^{2}\left(m^{2}\left(\lambda_{2}+\lambda_{3}\right)+M^{2}\left(\lambda_{1}+\lambda_{4}\right)\right)}{4\left(\lambda\left(\lambda_{2}+\lambda_{3}\right)-\left(\lambda_{1}+\lambda_{4}\right){}^{2}\right)}.\label{eq:0626-1-1}
\end{equation}

{\bf Case D: }$x\neq0$, $z\neq0$

This is a much more complicated case. Only $y$ is zero in this case.
The potential at $y=0$ is 
\begin{eqnarray}
V_{0}|_{y=0} & = & -\frac{m^{2}}{2}h^{2}+M^{2}(x^{2}+z^{2})+\frac{\lambda}{16}h^{4}+\lambda_{23}(x^{4}+z^{4})\nonumber \\
 &  & +2\lambda_{2}x^{2}z^{2}+\frac{h^{2}}{2}(\lambda_{14}x^{2}+\lambda_{1}z^{2}),\label{eq:t2-17}
\end{eqnarray}
where $\lambda_{23}\equiv\lambda_{2}+\lambda_{3}$, $\lambda_{14}\equiv\lambda_{1}+\lambda_{4}$.
Define 
\begin{equation}
Q\equiv\frac{1}{2}\left(\begin{array}{ccc}
\frac{\lambda}{4} & \lambda_{1}+\lambda_{4} & \lambda_{1}\\
\lambda_{14} & 4\lambda_{23} & 4\lambda_{2}\\
\lambda_{1} & 4\lambda_{2} & 4\lambda_{23}
\end{array}\right),\label{eq:t2-18}
\end{equation}
and 
\begin{equation}
b\equiv\left(\begin{array}{c}
m^{2}/2\\
-M^{2}\\
-M^{2}
\end{array}\right),\thinspace u\equiv\left(\begin{array}{c}
h^{2}\\
x^{2}\\
z^{2}
\end{array}\right),\label{eq:t2-19}
\end{equation}
then Eq. (\ref{eq:t2-17}) can be written as
\begin{equation}
V_{0}|_{y=0}=\frac{1}{2}u^{T}Qu-bu.\label{eq:t2-20}
\end{equation}
From $\partial V_{0}/\partial u=0$ we get $Qu-b=0$. So the minimum
should be at 
\begin{equation}
u=Q^{-1}b.\label{eq:t2-21}
\end{equation}
Here we need to compute the inverse of $Q$, 
\begin{equation}
Q^{-1}=\frac{1}{\det Q}\left(\begin{array}{ccc}
-4\lambda_{2}^{2}+4\lambda_{23}^{2} & \lambda_{1}\lambda_{2}-\lambda_{14}\lambda_{23} & \lambda_{2}\lambda_{14}-\lambda_{1}\lambda_{23}\\
\lambda_{1}\lambda_{2}-\lambda_{14}\lambda_{23} & \frac{1}{4}\left(-\lambda_{1}^{2}+\lambda\lambda_{23}\right) & \frac{1}{4}\left(-\lambda\lambda_{2}+\lambda_{1}\lambda_{14}\right)\\
\lambda_{2}\lambda_{14}-\lambda_{1}\lambda_{23} & \frac{1}{4}\left(-\lambda\lambda_{2}+\lambda_{1}\lambda_{14}\right) & \frac{1}{4}\left(-\lambda_{14}^{2}+\lambda\lambda_{23}\right)
\end{array}\right),\label{eq:t2-22}
\end{equation}
where 
\begin{equation}
\det Q=\frac{1}{2}(-\lambda\lambda_{2}^{2}+2\lambda_{1}\lambda_{2}\lambda_{14}-\lambda_{1}^{2}\lambda_{23}-\lambda_{14}^{2}\lambda_{23}+\lambda\lambda_{23}^{2}).\label{eq:t2-39}
\end{equation}
Combining the above results, the minimum is at 
\begin{equation}
h^{2}=\frac{2\lambda_{3}\left(2m^{2}\left(2\lambda_{2}+\lambda_{3}\right)+M^{2}\left(2\lambda_{1}+\lambda_{4}\right)\right)}{\lambda_{3}\left(-2\lambda_{1}^{2}+\lambda\left(2\lambda_{2}+\lambda_{3}\right)\right)-2\lambda_{1}\lambda_{3}\lambda_{4}-\left(\lambda_{2}+\lambda_{3}\right)\lambda_{4}^{2}},\label{eq:t2-23}
\end{equation}
\begin{equation}
x^{2}=\frac{\left(M^{2}\lambda_{1}+2m^{2}\lambda_{2}\right)\lambda_{4}+\lambda_{3}\left(M^{2}\lambda+2m^{2}\left(\lambda_{1}+\lambda_{4}\right)\right)}{4\lambda_{1}^{2}\lambda_{3}-2\lambda\lambda_{3}\left(2\lambda_{2}+\lambda_{3}\right)+4\lambda_{1}\lambda_{3}\lambda_{4}+2\left(\lambda_{2}+\lambda_{3}\right)\lambda_{4}^{2}},\label{eq:t2-24}
\end{equation}
\begin{equation}
z^{2}=\frac{\left(M^{2}\lambda+2m^{2}\lambda_{1}\right)\lambda_{3}-\lambda_{4}\left(2m^{2}\lambda_{2}+M^{2}\left(\lambda_{1}+\lambda_{4}\right)\right)}{4\lambda_{1}^{2}\lambda_{3}-2\lambda\lambda_{3}\left(2\lambda_{2}+\lambda_{3}\right)+4\lambda_{1}\lambda_{3}\lambda_{4}+2\left(\lambda_{2}+\lambda_{3}\right)\lambda_{4}^{2}}.\label{eq:t2-25}
\end{equation}
We call it the Type D solution. The potential value at this minimum
is
\begin{equation}
V_{{\rm min}}^{D}=\frac{4m^{4}\lambda_{3}^{2}-M^{4}\lambda_{4}^{2}+2\lambda_{3}\left(M^{4}\lambda+2m^{2}\left(2m^{2}\lambda_{2}+M^{2}\left(2\lambda_{1}+\lambda_{4}\right)\right)\right)}{8\lambda_{1}^{2}\lambda_{3}-4\lambda\lambda_{3}\left(2\lambda_{2}+\lambda_{3}\right)+8\lambda_{1}\lambda_{3}\lambda_{4}+4\left(\lambda_{2}+\lambda_{3}\right)\lambda_{4}^{2}}.\label{eq:t2-26}
\end{equation}

Then we consider  solutions of Eq. (\ref{eq:t2-13}) with nonzero
$y$. Again, there are many subcases depending on whether $h$, $x$,
or $z$ in the solution is zero. If they are all zero, then the potential
reduces to 
\begin{equation}
V_{0}|_{h=x=z=0}=M^{2}y^{2}+(\lambda_{2}+\frac{\lambda_{3}}{2})y^{4},\label{eq:t2-28}
\end{equation}
which is impossible to produce a nonzero minimum for $y$. So this
subcase is excluded. More generally, we can prove that there can not
be any local minima with $h=0$. If a local minimum is at $(h,x,y,z)=(0,x_{0},y_{0},z_{0})$
then we can assume $x_{0}\geq0$ without loss of generality. This
is because the potential
\begin{eqnarray}
V_{0}|_{h=0} & = & M^{2}(x^{2}+y^{2}+z^{2})+\lambda_{2}(x^{2}+y^{2}+z^{2})^{2}\nonumber \\
 &  & +\lambda_{3}[x^{4}+z^{4}+\frac{y^{4}}{2}+y^{2}x^{2}+y^{2}z^{2}+(x-z)^{2}]\label{eq:t2-27}
\end{eqnarray}
is invariant under the transformation $(x,z)\rightarrow(-x,-z)$.
If $x_{0}=0$ then $(0,x_{0},y_{0},z_{0})$ is not a local minimum,
because there is a direction in which the potential drops down. The
direction is $(0,x_{0},y_{0},z_{0})\rightarrow(0,x_{0},y_{0},z_{0}-z_{0}dt)$
where $dt$ stands for an infinitesimal step. If $x_{0}>0$ then there
is also a going-down direction, depending on whether $z_{0}$ is positive
or not. If $z_{0}\geq0$, then the direction is $(0,x_{0},y_{0},z_{0})\rightarrow(0,x_{0}-x_{0}dt,y_{0},z_{0}-x_{0}dt)$;
if $z_{0}<0$, it is $(0,x_{0},y_{0},z_{0})\rightarrow(0,x_{0}-x_{0}dt,y_{0},z_{0})$.
As one can check from the form given by Eq. (\ref{eq:t2-27}) that
the potential always goes down when it is moving from $(0,x_{0},y_{0},z_{0})$
in these directions. So $(0,x_{0},y_{0},z_{0})$ is impossible to
be a local minimum. 

For the other case that both $y$ and $h$ are nonzero, we conjecture
that no local minima exist in this case either. Although it is difficult
to analytically prove this, the conjecture can be verified numerically
by a random scan in the parameter space. Among $10^{5}$ randomly
generated samples, after numerical minimization, none of the minima
is found to have simultaneously nonzero $h$ and $y$.

In summary, there are only four types of local minima at the LO.
They are summarized in Tab.\,\ref{tab:All-local-minima}, where Type
A is the simplest solution, corresponding to the VEVs required to
generate tiny neutrino masses at the NLO. Type B and D break the electromagnetic
symmetry $U(1)_{{\rm em}}$ which must be avoided.  Type C respects
the $U(1)_{em}$ symmetry, but it generally leads to a large VEV of
$\Delta$ comparable to the VEV of $H$, which strongly violates the
custodial symmetry. Such a large VEV of $\Delta$ would be ruled out
by the $\rho$ parameter \cite{Agashe:2014kda}. Besides, it could
generate neutrino masses comparable to other fermion masses, which
also contradicts the experimental facts. Therefore, though there are
four possible types of minima, only Type A is allowed in the real
world. However Type B, C and D minima do exist in a certain region
of the parameter space of the Type II seesaw model. So it is possible
that the scalars may fall into some dangerous minima that would completely
invalidate the Type II seesaw model. We will see in the next subsection
 how likely this would happen.

\subsection{Numerical verification}

The result listed in Tab.\,\ref{tab:All-local-minima} can be verified
numerically by taking some random values for the potential parameters
and then using a minimization algorithm to find the  numerical minimum.
If the analytical result is correct, then the numerical minimum found
by computer should be identical to one of the four minima in Tab.\,\ref{tab:All-local-minima}
since they are the only four possible minima.

\begin{figure}
\centering

\includegraphics[width=8cm]{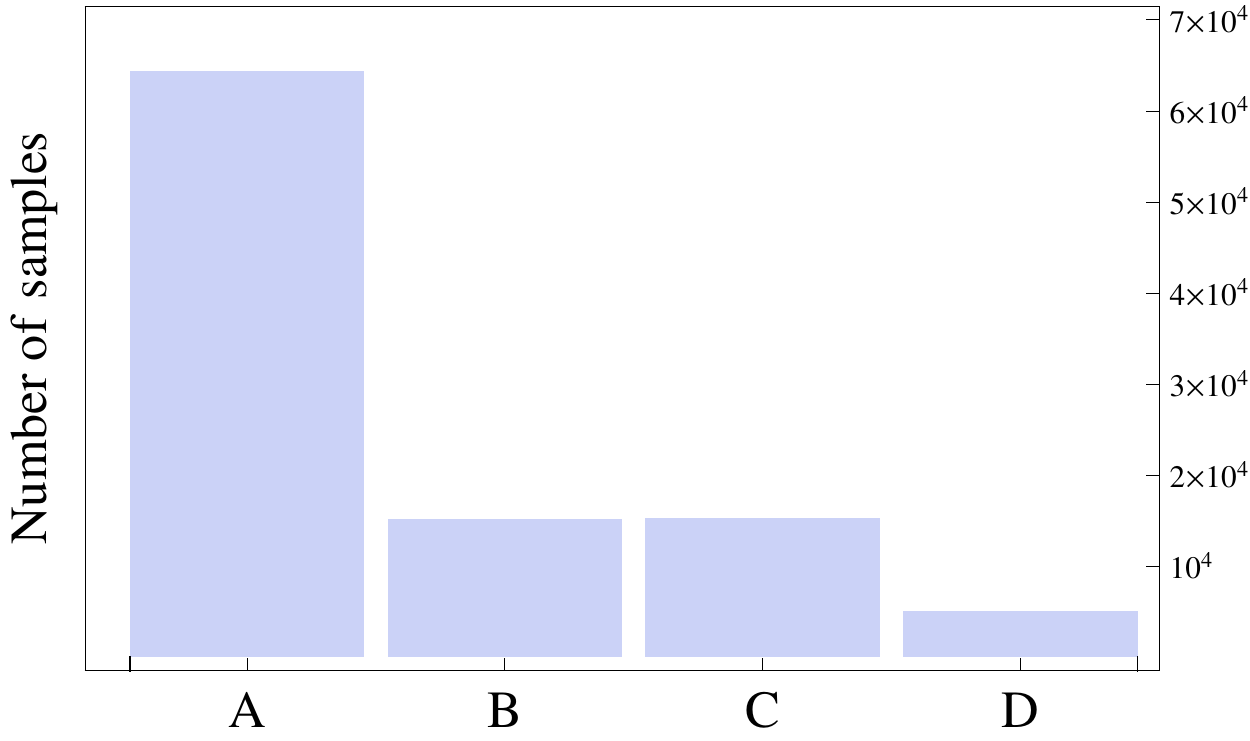}\caption{\label{fig:abcd}Distribution of the global minima of $10^{5}$ randomly
generated samples. There are only four possible types of minima A,
B, C and D (see Tab.\,\ref{tab:All-local-minima}). Type A is the
one conventionally adopted to generate tiny neutrino masses in the
model. The other types of minima  strongly contradict the experimental
facts. Nevertheless they are possible to emerge, as shown in this
figure,  depending on the potential parameters.}
\end{figure}

Furthermore, by repeating this process many times, we will statistically
obtain a distribution of these minima, to see how likely it is for
one of them to be the global minimum (the computer will always numerically
find the global minimum, to be explained later) if we randomly choose
a point in the parameter space. In Fig.\,\ref{fig:abcd} we show
such a distribution from $N=10^{5}$ samples. To generate the distribution,
we randomly select $10^{5}$ points in the parameter space of the
potential and numerically minimize the potential, count the number
of the points that lead to Type A, B, C or D minima respectively.
The result is 
\begin{equation}
(N_{A},N_{B},N_{C},N_{D})=(64419,15216,15304,5061),\label{eq:t2-29}
\end{equation}
where $N_{X}$ is the number corresponding to Type $X$ ($X=$ A,
B, C, D). 

Note that the sum of the numbers in (\ref{eq:t2-29}) equals to the
total number $N=10^{5}$ , i.e.
\begin{equation}
\sum_{X=A,B,C,D}N_{X}=N,\label{eq:t2-30}
\end{equation}
which implies that among the $10^{5}$ samples, there are not any
other new types of minima. So it numerically verifies the conclusion
that Type A, B, C and D are the only four possible  minima of the
potential. 

Besides, each of the four minima is possible to be the global minimum,
depending on the parameters of the potential. In other words, a general
potential of the Type II seesaw model does not necessarily lead to
the VEVs we want to generate neutrino masses. As previously mentioned,
only when the vacuum falls into the Type A minimum, it can be a successful
theory in accord with experiments. From Fig.\,\ref{fig:abcd} we
can see that Type A is the most likely to be the global minimum which
implies that those parameter configurations for Type A global minima
occupy the largest region in the parameter space. However, this is
not enough. We would like to know where the region is, i.e. when the
scalar fields will always get the VEVs we want. This will be studied
in the next section.

At the end of this section, we explain some technical details used
in the above numerical minimization. There have been many numerical
methods of multidimensional minimization, such as Newton's (or Quasi-Newton)
method, Nelder-Mead simplex algorithm \cite{NelderMead65}, differential
evolution \cite{Storn1997}, etc. So far for general multivariable
functions there have not been any numerical methods that can guarantee
to find the global minimum. All the known methods are based on iterative
algorithms and if the iteration converges at some point, it only means
the point is a local minimum. To find the global minimum, in principle
one can try many different initial points so that as many as possible
local minima are found. If enough initial points are searched, the
global minimum would be found. However, there is no criterion for
how many initial points should be selected and too many initial points
could be computationally very expensive.

In this work, to verify that there are only four types of minima,
we only need an algorithm to find a local minimum. Whether it is global
or not  does not concern us at this stage. When the process is done
for many ($10^{5}$) randomly generated potentials and the minima
(the initial point is also randomly set for each potential) found
are always in the four types, then the conclusion is verified at a
very high confidence level. Going to the next stage, we want to make
the program always find the global minimum. This can be done based
on the verified conclusion that there are only four types of minima,
which have been analytically computed. So we only need to compare
the four known local minima to get the global minimum. Therefore,
improved by the analytical result, the program can always find the
global minimum.

Finally, there is something non-trivial about the parameter space.
The parameters of the potential can not be arbitrary real numbers
because the potential must be bounded from below (BFB). In Refs.\,\cite{Arhrib:2011uy,Bonilla:2015eha}
the BFB constraint has been studied and the necessary and sufficient
constraint has a slightly complicated form \cite{Bonilla:2015eha}.
Besides, there are also some unitarity constraints which require those
$\lambda$'s ($\lambda$ and $\lambda_{1,2,3,4}$) and their combinations
not larger than some certain values, typically $4\pi$. But this is
not relevant here because an overall rescaling of all the $\lambda$'s
can not change the properties of the minima. So in this work, we simply
take all the $\lambda$'s in the interval from $-1$ to $1$, and
then reject those selections which violate the BFB constraint from
Ref.\,\cite{Bonilla:2015eha}. As for $m$ and $M$ in the potential,
since their scale is also not important, we set $M=1$ and $m\in(0,10)$.
The BFB condition from Ref.\,\cite{Bonilla:2015eha} is actually
checked again in the numerical minimization because the iteration
would not converge if the potential is not bounded from below. So
this verifies that the BFB constraint from Ref.\,\cite{Bonilla:2015eha}
is at least sufficient.

\section{\label{sec:From-local-to}From local to global}

As we have seen in Fig.\,\ref{fig:abcd}, the potential of the Type
II seesaw model does not necessarily lead to the VEVs corresponding
to the Type A minimum. If the Type A minimum is not the deepest (i.e.
the global minimum), then there is a danger that the early universe
might fall into a deeper minimum by quantum tunneling, which could
be very different from the universe we are living in  today. So we
would like to know when the Type A minimum is global.

\begin{figure}
\centering

\includegraphics[width=7.5cm]{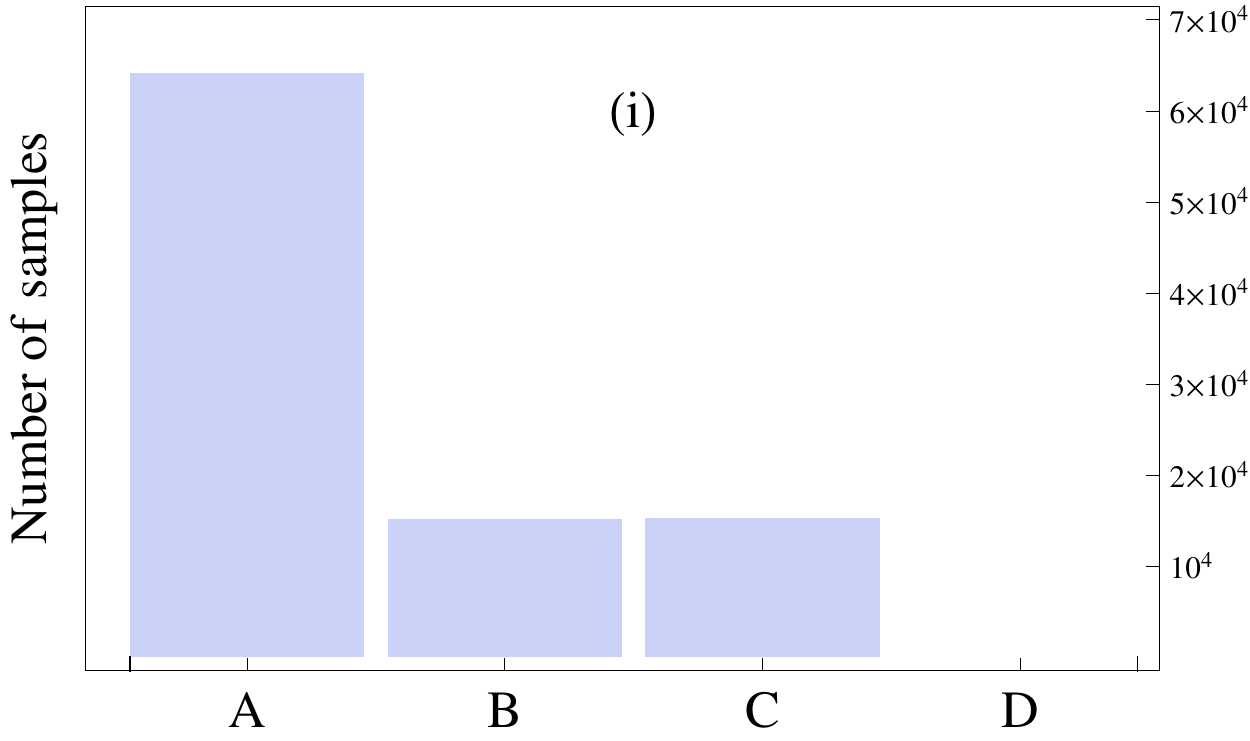}\hspace{1cm}\includegraphics[width=7.5cm]{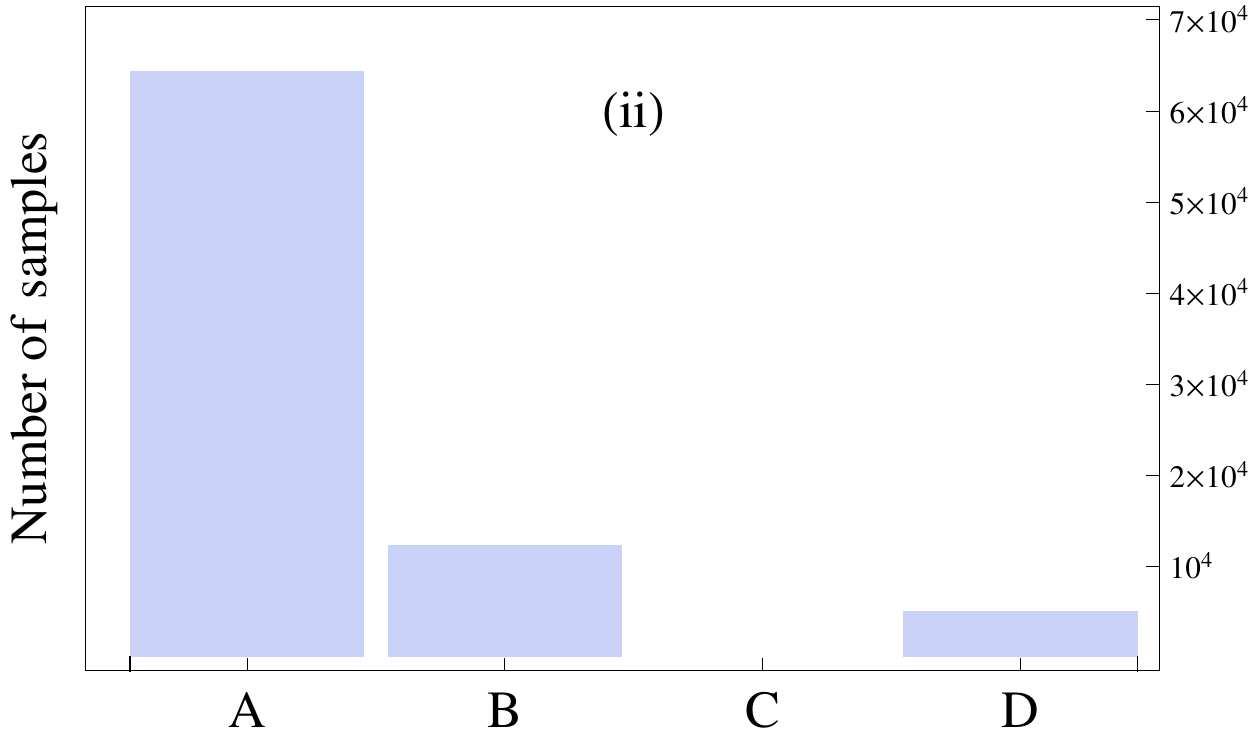}

\includegraphics[width=7.5cm]{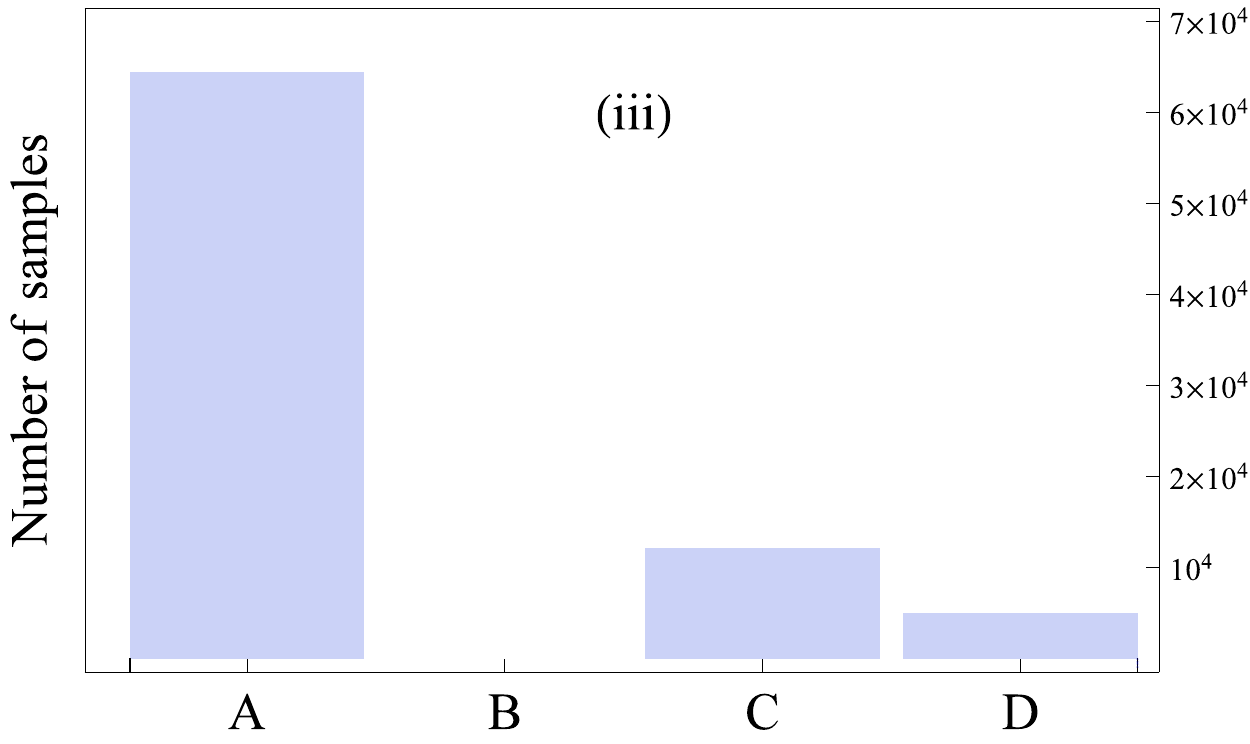}\hspace{1cm}\includegraphics[width=7.5cm]{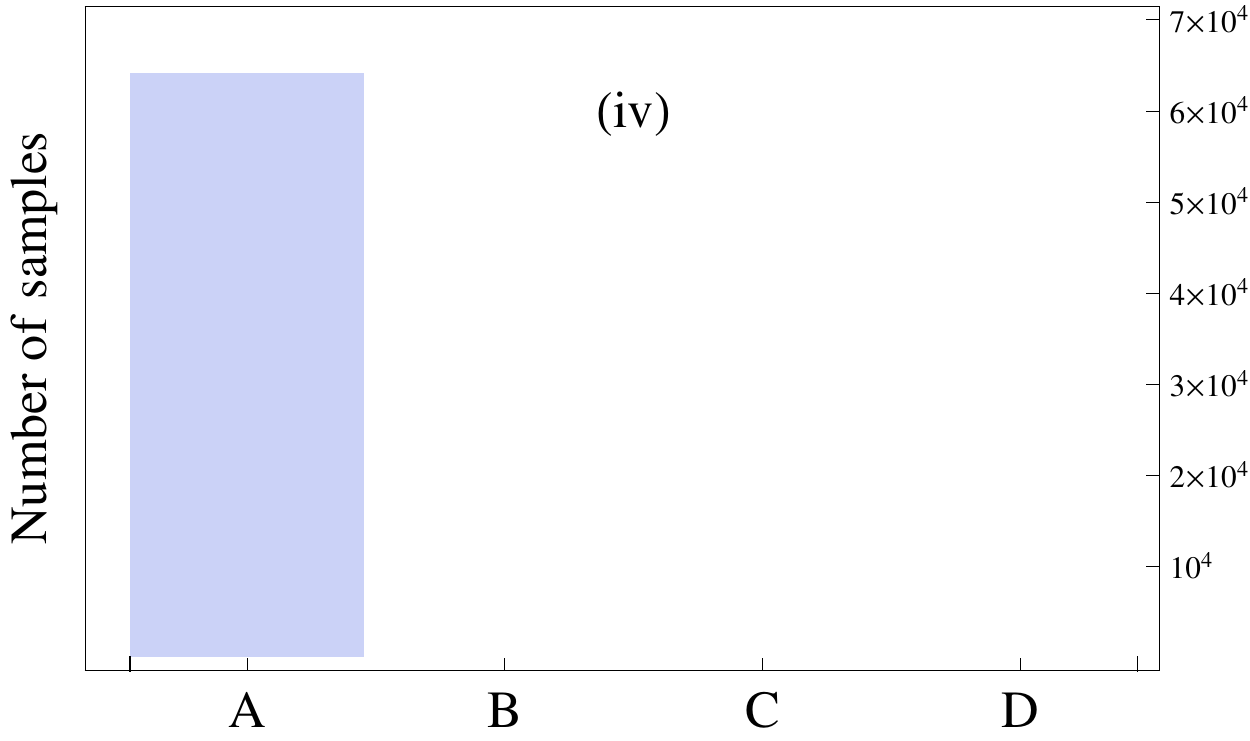}\caption{\label{fig:togloabl}Similar plots to Fig.\,\ref{fig:abcd} but the
potential parameters are constrained by conditions B, C and D given
by (\ref{eq:t2-31})-(\ref{eq:t2-33}). Partially adding these constraints
will avoid Type D, C, or B minima being global, shown in plots (i),
(ii) or (iii) respectively. Combining all of them leads to plot (iv)
where all the global minima are of Type A.}
\end{figure}

Since there are only four types of minima,  we can compare the Type
A minimum with other minima by their potential values. If the Type
A minimum is deeper than all the other minima, then it must be the
global minimum. The result of such an comparison (analyses given afterwards)
is that, to make the Type A minimum global, we need some constraints
on the potential parameters, given by 
\begin{equation}
{\rm condition\thinspace B}:\thinspace\lambda+2\eta^{2}\lambda_{1}\geq0\quad{\rm or}\quad\lambda_{1}+2\eta^{2}\left(\lambda_{2}+\lambda_{3}\right)\leq0,\label{eq:t2-31}
\end{equation}
\begin{equation}
{\rm condition\thinspace C}:\thinspace\lambda+2\eta^{2}\left(\lambda_{1}+\lambda_{4}\right)\geq0\quad{\rm or}\quad\lambda_{1}+2\eta^{2}\left(\lambda_{2}+\lambda_{3}\right)+\lambda_{4}\leq0,\label{eq:t2-32}
\end{equation}
\begin{eqnarray}
{\rm condition\thinspace D}: &  & 4\lambda_{3}\left(2\lambda_{1}+2\eta^{2}\left(2\lambda_{2}+\lambda_{3}\right)+\lambda_{4}\right)\leq0\nonumber \\
 & {\rm or}\quad & \left(\lambda+2\eta^{2}\lambda_{1}\right)\lambda_{3}\geq\lambda_{4}\left(\lambda_{1}+2\eta^{2}\lambda_{2}+\lambda_{4}\right)\nonumber \\
 & {\rm or}\quad & \left(\lambda_{1}+2\eta^{2}\lambda_{2}\right)\lambda_{4}+\lambda_{3}\left(\lambda+2\eta^{2}\left(\lambda_{1}+\lambda_{4}\right)\right)\geq0,\label{eq:t2-33}
\end{eqnarray}
where $\eta\equiv m/M$ and condition X comes from the comparison
of Type A with Type X minima. When the parameters satisfy all these
conditions, then the Type A minimum must be global, as shown in Fig.\,\ref{fig:togloabl}.
Similar to Fig.\,\ref{fig:abcd}, in Fig.\,\ref{fig:togloabl} we
first randomly generated $10^{5}$ sets of parameters allowed by the
BFB condition. The difference is that before minimization, we drop
some of the samples that violate the above conditions.

For example, if we drop the samples that violate condition D, then
after numerical minimization we find all the minima are of Type A,
B or C, i.e. none of them falls into Type D, as shown in plot (i)
of Fig.\,\ref{fig:togloabl}. If we drop those samples that violate
condition C (or B) but not condition D, then we get plot (ii) {[}or
plot (iii){]}. If all these constraint are applied, i.e. we remove
those samples that violate any of the three conditions,  then the
remaining samples always fall into Type A minima, shown in plot (iv).
Therefore, if the parameters satisfy all the conditions given by (\ref{eq:t2-31})-(\ref{eq:t2-33}),
the global minimum of such a potential must be of Type A.

Next we shall derive the conditions (\ref{eq:t2-31})-(\ref{eq:t2-33})
from the comparison. Although there are analytic expressions for the
four solutions in Tab.\,\ref{tab:All-local-minima}, for a certain
set of parameters some of them may not be real solutions because $h^{2}$,
$x^{2}$, $y^{2}$ or $z^{2}$ from these solutions could be negative,
implying $h$, $x$, $y$ or $z$ would be imaginary at these points.
So before comparing the Type A minimum with other minima, we first
need to make it clear when they are real solutions.

Type A is fine in any case since $\frac{4m^{2}}{\lambda}>0$. Here
$\lambda>0$ is required by the BFB condition \cite{Arhrib:2011uy,Bonilla:2015eha}.
Type B is more complicated, because whether $h^{2}$ and $z^{2}$
are positive or not depends on the denominator $\lambda_{1}^{2}-\lambda\left(\lambda_{2}+\lambda_{3}\right)$.
But the BFB condition only requires \cite{Arhrib:2011uy} 
\begin{equation}
\lambda_{1}\geq-\sqrt{\lambda(\lambda_{2}+\lambda_{3})},\label{eq:t2-34}
\end{equation}
which can not tell us whether $\lambda_{1}^{2}-\lambda\left(\lambda_{2}+\lambda_{3}\right)$
is negative or positive. Actually the potential is bounded from below
if and only if the matrix of quartic couplings is copositive \cite{Kannike:2012pe,Chakrabortty:2013mha}.
In the case of Type B, the matrix is 
\begin{equation}
Q_{B}=\left(\begin{array}{cc}
\frac{\lambda}{16} & \frac{\lambda_{1}}{4}\\
\frac{\lambda_{1}}{4} & \lambda_{2}+\lambda_{3}
\end{array}\right).\label{eq:t2-35}
\end{equation}
If $Q_{B}$ is positive definite, which requires that 
\begin{equation}
\det Q_{B}=-\frac{\lambda_{1}^{2}-\lambda(\lambda_{2}+\lambda_{3})}{16}>0,\label{eq:t2-36}
\end{equation}
then for any $2\times1$ vector $u$ we always have $u^{T}Q_{B}u>0$.
In this case, the potential is always bounded from below. However,
$Q_{B}$ being positive definite is a sufficient but not necessary
condition of BFB. Since $u=(h^{2},\thinspace z^{2})^{T}$ only contains
positive components, if all the entries of $Q_{B}$ are positive,
then we still have $u^{T}Q_{B}u>0$. So the potential is bounded from
below either if $Q_{B}$ is positive definite, or if all the entries
of $Q_{B}$ are positive. It can be checked that this is the necessary
and sufficient condition \cite{Kannike:2012pe}. But if all the entries
are positive, then the numerators in Eqs.(\ref{eq:t2-15}) and (\ref{eq:t2-16})
should be positive , i.e. $M^{2}\lambda_{1}+2m^{2}\left(\lambda_{2}+\lambda_{3}\right)>0$
and $M^{2}\lambda+2m^{2}\lambda_{1}>0$, which implies $h^{2}z^{2}<0$
and thus the Type B solution can not be real, no matter $\lambda_{1}^{2}-\lambda(\lambda_{2}+\lambda_{3})>0$
or not. So only the remaining case that $Q_{B}$ is positive definite
concerns us. For positive definite $Q_{B}$, since $\lambda_{1}^{2}-\lambda(\lambda_{2}+\lambda_{3})<0$,
if $M^{2}\lambda_{1}+2m^{2}\left(\lambda_{2}+\lambda_{3}\right)>0$
and $M^{2}\lambda+2m^{2}\lambda_{1}<0$, then we have positive $h^{2}$
and $z^{2}$ for the Type B solution. Note that $\lambda_{1}$ can
not be positive if the two numerators have opposite signs, thus we
can draw the conclusion that as long as the potential is bounded from
below, the Type B solution is real if and only if $M^{2}\lambda_{1}+2m^{2}\left(\lambda_{2}+\lambda_{3}\right)>0$
and $M^{2}\lambda+2m^{2}\lambda_{1}<0$. 

If this condition is satisfied, then we need to compare the potential
values at the Type A and B minima, denoted as $V_{0}^{(A)}$ and $V_{0}^{(B)}$
respectively. By straightforward calculation, the difference $V_{0}^{(A)}-V_{0}^{(B)}$
is 
\begin{equation}
V_{0}^{(A)}-V_{0}^{(B)}=\frac{M^{4}\left(\lambda+2m^{2}\lambda_{1}\right){}^{2}}{4\lambda\left(-\lambda_{1}^{2}+\lambda(\lambda_{2}+\lambda_{3})\right)},\label{eq:t2-37}
\end{equation}
which is positive if $\lambda>0$ and $-\lambda_{1}^{2}+\lambda(\lambda_{2}+\lambda_{3})>0$.
$\lambda>0$ can be guaranteed by the BFB condition while $-\lambda_{1}^{2}+\lambda(\lambda_{2}+\lambda_{3})>0$
is satisfied as long as the Type B solution is real, explained as
follows. As just mentioned, if Type B is real, then it excludes the
possibility of positive $\lambda_{1}$, otherwise $h^{2}z^{2}$ would
be negative. Then from (\ref{eq:t2-34}) we have $-\lambda_{1}^{2}+\lambda(\lambda_{2}+\lambda_{3})>0$.
So if the Type B solution is real then the potential value at the
Type B minimum must be lower than Type A. To avoid that, we need to
add some constraints so that the real Type B solution can not exist.
Therefore either the condition $M^{2}\lambda_{1}+2m^{2}\left(\lambda_{2}+\lambda_{3}\right)>0$
or $M^{2}\lambda+2m^{2}\lambda_{1}<0$ should be violated, which gives
 the result given by (\ref{eq:t2-31}).

Next let us turn to the Type C solution. The analysis is similar and
we find that the Type C solution is real if and only if $2m^{2}\left(\lambda_{2}+\lambda_{3}\right)+M^{2}\left(\lambda_{1}+\lambda_{4}\right)>0$
and $M^{2}\lambda+2m^{2}\left(\lambda_{1}+\lambda_{4}\right)<0$.
The difference $V_{0}^{(A)}-V_{0}^{(C)}$ where $V_{0}^{(C)}$ is
the potential value at the Type C minimum is
\begin{equation}
V_{0}^{(A)}-V_{0}^{(C)}=\frac{M^{4}\left(\lambda+2m^{2}\left(\lambda_{1}+\lambda_{4}\right)\right){}^{2}}{4\lambda\left(\lambda\left(\lambda_{2}+\lambda_{3}\right)-\left(\lambda_{1}+\lambda_{4}\right){}^{2}\right)}.\label{eq:t2-38}
\end{equation}
Again, this implies that to avoid a deeper minimum at Type C, we need
to break the existence condition of the real Type C solution, which
leads to the result given by (\ref{eq:t2-32}).

Finally there is the Type D solution. Though it has the most complicated
form, we find that the analysis is still similar. The difference $V_{0}^{(A)}-V_{0}^{(D)}$
can be written as 
\begin{equation}
V_{0}^{(A)}-V_{0}^{(D)}=\frac{M^{4}\left(\lambda+2m^{2}\lambda_{1}\right){}^{2}}{4\lambda\left(\lambda\left(\lambda_{2}+\lambda_{3}\right)-\lambda_{1}^{2}\right)}+\frac{M^{4}\left(\left(\lambda_{1}+2m^{2}\lambda_{2}\right)\lambda_{4}+\lambda_{3}\left(\lambda+2m^{2}\left(\lambda_{1}+\lambda_{4}\right)\right)\right){}^{2}}{8\det Q\left(\lambda\left(\lambda_{2}+\lambda_{3}\right)-\lambda_{1}^{2}\right)},\label{eq:t2-40}
\end{equation}
where $\det Q$ has been defined in Eq.(\ref{eq:t2-39}). Again, the
denominators are positive because those determinants are positive.
The numerators are positive because they are in squared forms. So
a similar analysis gives the result in (\ref{eq:t2-33}).

\section{\label{sec:NLO}To The NLO}

The above analysis is only for the LO, i.e. assuming $\mu=0$. However,
in the limit of zero $\mu$, the model can not account for tiny neutrino
masses. In this section, we will introduce a  perturbatively small
but nonzero $\mu$ into our calculation, to see how those local minima
are modified by the NLO correction. 

First we would like to discuss in general the perturbative calculation
near minima. Consider a general potential $V(\phi)$ which is a function
of multifields $\phi=(\phi_{1},\phi_{2},\cdots)$ and can be written
as 
\begin{equation}
V(\phi)=V_{0}(\phi)+\delta V(\phi),\label{eq:t2-41}
\end{equation}
where $V_{0}$ and $\delta V$ are the LO and NLO terms. Assume that
$V_{0}$ has a minimum at $\phi^{(0)}$, i.e. 
\begin{equation}
\left.\frac{\partial V_{0}}{\partial\phi_{i}}\right|_{\phi=\phi^{(0)}}=0,\thinspace\thinspace(i=1,2,3\cdots),\label{eq:t2-42}
\end{equation}
then the corresponding minimum of $V$, computed at the NLO should
be approximately at
\begin{equation}
\phi^{(1)}=\phi^{(0)}+\delta\phi,\label{eq:t2-43}
\end{equation}
where $\delta\phi$ can be determined from 
\begin{equation}
\delta\phi=-\left.\left[\frac{\partial^{2}V_{0}}{\partial\phi_{i}\partial\phi_{j}}\right]^{-1}\frac{\partial\delta V}{\partial\phi}\right|_{\phi=\phi^{(0)}}.\label{eq:t2-44}
\end{equation}
The potential value at this minimum is
\begin{equation}
V(\phi^{(1)})=V_{0}(\phi^{(0)})+\delta V(\phi^{(0)})+O(\delta\phi^{2}).\label{eq:t2-45}
\end{equation}
Here we briefly derive Eqs.\,(\ref{eq:t2-44}) and (\ref{eq:t2-45}).
Consider the equation $\partial V/\partial\phi=0$, which is 
\begin{equation}
\left.\frac{\partial V_{0}(\phi)}{\partial\phi_{i}}+\frac{\partial\delta V(\phi)}{\partial\phi_{i}}\right|_{\phi=\phi^{(0)}+\delta\phi}=0.\label{eq:t2-47}
\end{equation}
The first term can be expanded as 
\begin{equation}
\left.\frac{\partial V_{0}(\phi)}{\partial\phi_{i}}\right|_{\phi=\phi^{(0)}+\delta\phi}=\left.\frac{\partial V_{0}(\phi)}{\partial\phi_{i}}\right|_{\phi=\phi^{(0)}}+\delta\phi_{j}\left.\frac{\partial^{2}V_{0}}{\partial\phi_{i}\partial\phi_{j}}\right|_{\phi=\phi^{(0)}}+O(\delta\phi^{2}),\label{eq:t2-46}
\end{equation}
while the second term of Eq.\,(\ref{eq:t2-47}) is
\begin{equation}
\left.\frac{\partial\delta V(\phi)}{\partial\phi_{i}}\right|_{\phi=\phi^{(0)}+\delta\phi}=\left.\frac{\partial\delta V(\phi)}{\partial\phi_{i}}\right|_{\phi=\phi^{(0)}}+O(\delta\phi^{2}).\label{eq:t2-48}
\end{equation}
Using Eq.\,(\ref{eq:t2-42}) we write Eq.\,(\ref{eq:t2-47}) as
\[
\delta\phi_{j}\left.\frac{\partial^{2}V_{0}}{\partial\phi_{i}\partial\phi_{j}}\right|_{\phi=\phi^{(0)}}+\left.\frac{\partial\delta V(\phi)}{\partial\phi_{i}}\right|_{\phi=\phi^{(0)}}+O(\delta\phi^{2})=0.
\]
which is a linear equation of $\delta\phi$. Solving the equation,
we get  Eq.\,(\ref{eq:t2-44}). Eq.\,(\ref{eq:t2-45}) can be obtained
by directly expanding the potential value near the LO minimum. Because
$\partial V_{0}/\partial\phi$ at the LO minimum is zero and $\delta V(\phi^{(1)})=\delta V(\phi^{(0)})+O(\delta\phi^{2})$,
we simply get the result in Eq.\,(\ref{eq:t2-45}).

\begin{table}
\centering

\begin{tabular}{|c|c|c|c|c|}
\hline 
corrections & Type A & Type B & Type C & Type D\tabularnewline
\hline 
\rule[-5ex]{0pt}{12ex}   $\delta h/\mu$ & 0 & 0 & $\frac{\sqrt{-C_{1}C_{2}}\left(2C_{2}\left(\lambda_{2}+\lambda_{3}\right)+C_{1}\left(\lambda_{1}+\lambda_{4}\right)\right)}{C_{d}C_{1}C_{2}}$ & $\frac{\sqrt{-D_{1}D_{2}}\lambda_{3}\left(D_{1}\lambda_{1}+10D_{2}\lambda_{2}+2D_{3}\lambda_{2}-4D_{3}\lambda_{3}+3D_{1}\lambda_{4}\right)}{\sqrt{2}D_{d}D_{1}D_{2}}$\tabularnewline
\rule[-5ex]{0pt}{10ex}  $\delta x/\mu$ & $\frac{2m^{2}}{C_{2}}$ & $\frac{B_{1}}{B_{2}\lambda_{3}+B_{1}\lambda_{4}}$ & $-\frac{m^{2}}{C_{2}}-\frac{3\left(\lambda_{1}+\lambda_{4}\right)}{2C_{d}}$ & $\frac{D_{1}\lambda_{1}^{2}-\lambda D_{1}\left(\lambda_{2}+\lambda_{3}\right)-4D_{2}\left(\lambda_{1}\lambda_{3}+\lambda_{2}\lambda_{4}+\lambda_{3}\lambda_{4}\right)}{4D_{2}D_{d}}$\tabularnewline
$\delta y/\mu$ & 0 & 0 & 0 & 0\tabularnewline
\rule[-5ex]{0pt}{10ex}  $\delta z/\mu$ & 0 & 0 & 0 & $\frac{\sqrt{-D_{2}D_{3}}\left(-D_{1}\lambda_{1}^{2}+\lambda D_{1}\lambda_{2}+4D_{3}\lambda_{1}\lambda_{3}-3D_{1}\lambda_{1}\lambda_{4}+4D_{2}\lambda_{2}\lambda_{4}\right)}{4D_{d}D_{2}D_{3}}$\tabularnewline
\rule[-5ex]{0pt}{10ex}  $\delta V_{{\rm min}}/\mu$ & 0 & 0 & $-\sqrt{\frac{-C_{2}}{C_{d}}}\frac{\sqrt{2}C_{1}}{C_{d}}$ & $-\sqrt{\frac{-D_{2}}{D_{d}}}\frac{D_{1}}{\sqrt{2}D_{d}}$\tabularnewline
\hline 
\end{tabular}

\caption{\label{tab:NLO-corrections}NLO corrections of the solutions listed
in Tab.\,\ref{tab:All-local-minima}.}
\end{table}

With Eqs.\,(\ref{eq:t2-44}) and (\ref{eq:t2-45}), it is quite straightforward
to compute the NLO corrections. The result is listed in Tab.\,\ref{tab:NLO-corrections},
where $B_{1,2,d}$, $C_{1,2,d}$ and $D_{1,2,3,d}$ are defined below,
\begin{eqnarray}
 &  & B_{1}=M^{2}\lambda_{1}+2m^{2}\left(\lambda_{2}+\lambda_{3}\right),\label{eq:t2-49}\\
 &  & B_{2}=M^{2}\lambda+2m^{2}\lambda_{1},\label{eq:t2-50}\\
 &  & B_{d}=\lambda\left(\lambda_{2}+\lambda_{3}\right)-\lambda_{1}^{2},\label{eq:t2-51}\\
 &  & C_{1}=2m^{2}\left(\lambda_{2}+\lambda_{3}\right)+M^{2}\left(\lambda_{1}+\lambda_{4}\right),\label{eq:t2-52}\\
 &  & C_{2}=M^{2}\lambda+2m^{2}\left(\lambda_{1}+\lambda_{4}\right),\label{eq:t2-53}\\
 &  & C_{d}=\lambda\left(\lambda_{2}+\lambda_{3}\right)-\left(\lambda_{1}+\lambda_{4}\right){}^{2},\label{eq:t2-54}\\
 &  & D_{1}=2\lambda_{3}\left(2m^{2}\left(2\lambda_{2}+\lambda_{3}\right)+M^{2}\left(2\lambda_{1}+\lambda_{4}\right)\right),\label{eq:t2-55}\\
 &  & D_{2}=\left(M^{2}\lambda_{1}+2m^{2}\lambda_{2}\right)\lambda_{4}+\lambda_{3}\left(M^{2}\lambda+2m^{2}\left(\lambda_{1}+\lambda_{4}\right)\right),\label{eq:t2-56}\\
 &  & D_{3}=\lambda_{4}\left(2m^{2}\lambda_{2}+M^{2}\left(\lambda_{1}+\lambda_{4}\right)\right)-\left(M^{2}\lambda+2m^{2}\lambda_{1}\right)\lambda_{3},\label{eq:t2-57}\\
 &  & D_{d}=\lambda_{3}\left(-2\lambda_{1}^{2}+\lambda\left(2\lambda_{2}+\lambda_{3}\right)\right)-2\lambda_{1}\lambda_{3}\lambda_{4}-\left(\lambda_{2}+\lambda_{3}\right)\lambda_{4}^{2}.\label{eq:t2-58}
\end{eqnarray}

An interesting point is that even in the NLO correction, $y$ still
remains zero. That is to say, the diagonal entries of $\Delta$ always
acquire zero VEVs. Besides, the NLO correction to  $V_{{\rm min}}$
is zero for Type A and B minima. The reason is that the NLO correction
to $V_{{\rm min}}$ is proportional to the value of $x^{2}$ at LO,
which is zero for the Type A and B minima (cf. Tab.\,\ref{tab:All-local-minima}).

In the LO calculation, we have obtained the condition of the Type
A minimum being the global minimum. So we would like to know the corresponding
condition at the NLO. The method is the same as what we have done
in the LO analysis. Because we have the analytic expressions for all
the minima at the NLO, one may simply compare the four types of minima
at the NLO to find the global minimum. However, we find that the explicit
expression of the condition to keep the Type A minimum global at the
NLO is very complicated, since the explicit forms of the NLO solutions
are already very complicated, as one can see from Eqs.\,(\ref{eq:t2-49})-(\ref{eq:t2-58})
and Tab.\,\ref{tab:NLO-corrections}. So in practical use,  to check
if the potential has the Type A global minimum for a set of numerically
given parameters, we would recommend evaluating the expressions in
Tab.\,\ref{tab:All-local-minima} and Tab.\,\ref{tab:NLO-corrections}
directly and comparing the four minima, instead of using more complicated
expressions derived from them.

\section{\label{sec:Prediction}Predictions of the mass splitting}

After spontaneous symmetry breaking, there are five massive scalar
fields in the Type II seesaw model, including three neutral fields
$(h^{0},\thinspace H^{0},\thinspace A^{0})$, one singly-changed field
$H^{\pm}$ and one doubly-changed field $H^{\pm\pm}$. Apart from
 the five scalar masses, the vacuum expectation values $(v_{H},\thinspace v_{\Delta})$
and the mixing angle $\alpha$ between $h^{0}$ and $H^{0}$ are also
physical observables. The 8 physical observables can be used to reconstruct
the parameters in the potential, since the potential only has 8 parameters
$(m^{2},\thinspace M^{2},\thinspace\mu,\thinspace\lambda,\thinspace\lambda_{1,2,3,4})$.
For explicit expressions, see Ref.\,\cite{Arhrib:2011uy}. Conversely,
the 8 physical observables can also be expressed in terms of the 8
potential parameters, which enable us to convert constraints on the
potential parameters to physical predictions.

For simplicity, we only focus on the predictions of the scalar mass
spectrum. In the Type II seesaw model, usually $h^{0}$ is considered
as the Higgs boson with mass $m_{h^{0}}=125$ GeV that has been discovered
at the LHC since 2012 \cite{Aad:2012tfa,Chatrchyan:2012xdj}. The
other massive scalar bosons are assumed to be much heavier than $h^{0}$.
At the LO, their masses are given by \cite{Arhrib:2011uy} 
\begin{equation}
m_{H^{\pm}}^{2}=M^{2}+\left(2\lambda_{1}+\lambda_{4}\right)\frac{v_{H}^{2}}{4}+{\cal O}(\mu^{2}),\thinspace\thinspace m_{H^{\pm\pm}}^{2}=M^{2}+\lambda_{1}\frac{v_{H}^{2}}{2}+{\cal O}(\mu^{2}),\label{eq:t2-65}
\end{equation}
\begin{equation}
m_{A}^{2}=M^{2}+\left(\lambda_{1}+\lambda_{4}\right)\frac{v_{H}^{2}}{2}+{\cal O}(\mu^{2}),\thinspace\thinspace m_{H^{0}}^{2}=M^{2}+\left(\lambda_{1}+\lambda_{4}\right)\frac{v_{H}^{2}}{2}+{\cal O}(\mu^{2}),\label{eq:t2-66}
\end{equation}
which implies that if the Type II seesaw scale $M$ is much higher
than the electroweak scale $v_{H}$, then all the four bosons should
have masses approximately equal to $M$. However, there are still
mass splittings among them. Defining
\begin{equation}
\Delta m^{2}\equiv m_{H^{0}}^{2}-m_{H^{\pm\pm}}^{2},\label{eq:t2-67}
\end{equation}
we find that 
\begin{equation}
\Delta m^{2}\approx m_{A}^{2}-m_{H^{\pm\pm}}^{2}\approx2(m_{H^{\pm}}^{2}-m_{H^{\pm\pm}}^{2})\approx\frac{\lambda_{4}}{2}v_{H}^{2}.\label{eq:t2-68}
\end{equation}
There is yet another much smaller mass squared difference, 
\begin{equation}
\delta m^{2}\equiv m_{A}^{2}+m_{H^{\pm\pm}}^{2}-2m_{H^{\pm}}^{2}\approx(\lambda_{4}-\lambda_{3})v_{\Delta}^{2}.\label{eq:t2-69}
\end{equation}

\begin{figure}
\centering

\includegraphics[width=8cm]{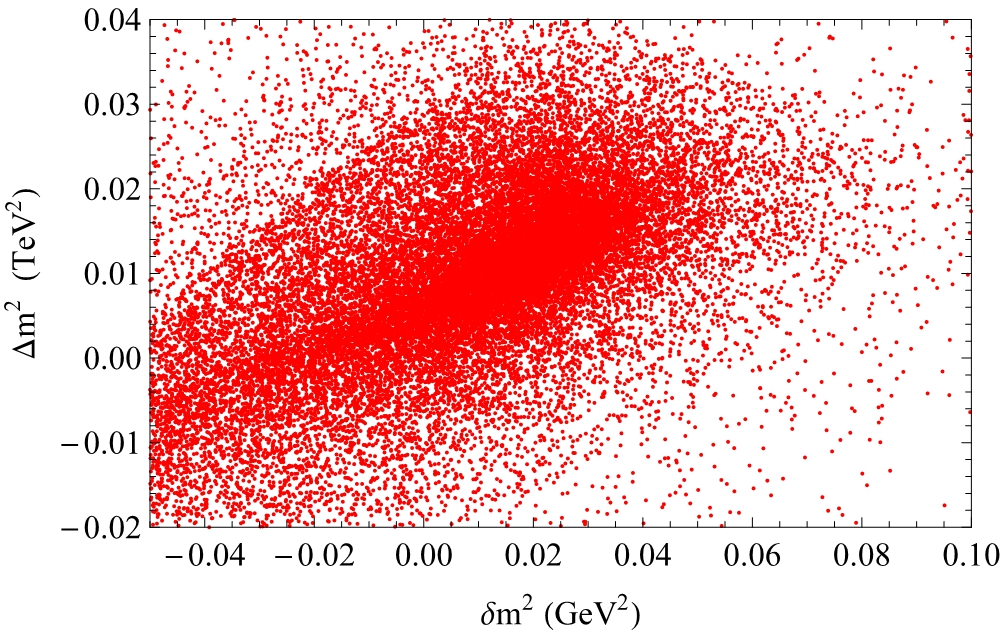}\,\includegraphics[width=8cm]{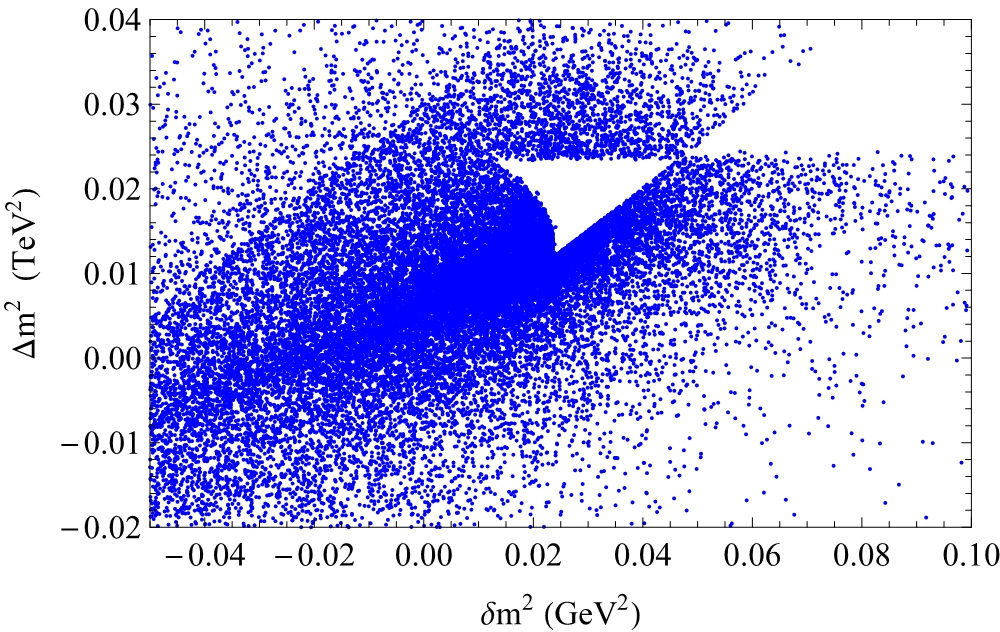}\caption{\label{fig:blue-red}The allowed region of the mass splitting parameters
$\Delta m^{2}\equiv m_{H^{0}}^{2}-m_{H^{\pm\pm}}^{2}$ and $\delta m^{2}\equiv m_{A}^{2}+m_{H^{\pm\pm}}^{2}-2m_{H^{\pm}}^{2}$
with (blue) or without (red) the requirement of the Type A vacuum
being the deepest.}
\end{figure}

If the potential is required to have the Type A global minimum, then
there may be additional constraints on $\delta m^{2}$ and $\Delta m^{2}$.
We use random search to find the constraints numerically, as shown
in Fig.\,\ref{fig:blue-red}. In the left panel, we randomly generate
$4\times10^{4}$ samples with all the potential parameters subject
to the BFB constraint from Ref.\,\cite{Bonilla:2015eha} and compute
the corresponding values of $\delta m^{2}$ and $\Delta m^{2}$ while
in the right panel, the samples are further constrained by Eqs.\,(\ref{eq:t2-31})-(\ref{eq:t2-33}).
To make it realistic, we fix the Higgs mass $m_{h^{0}}$ at 125 GeV,
and $v=\sqrt{v_{H}^{2}+v_{\Delta}^{2}}$ at 246 GeV. Also we assume
$m_{H^{0}}$ should be heavy ($m_{H^{0}}=$1 TeV) while $v_{\Delta}$
and $\alpha$ should be small ($v_{\Delta}=0.245$ GeV, $\tan2\alpha=4\times10^{-3}$).
Thus, the potential parameters are also subject to these constraints.
As we can see, there is a difference between the left and right panels
in Fig.\,\ref{fig:blue-red}: in some region of the right panel blue
points do not appear, which implies that  the corresponding values
of $(\delta m^{2},\thinspace\Delta m^{2})$ are not allowed if the
Type A minimum is global.

\section{\label{sec:Discussion-and-Conclusion}Conclusion}

In the Type II seesaw model, the vacuum could be of four different
types,  namely Type A, B, C and D. The analytic expressions of the
VEVs are listed in Tab.\,\ref{tab:All-local-minima} for the LO
and Tab.\,\ref{tab:NLO-corrections} for the NLO. Among the four
vacua only one of them, the Type A vacuum, is allowed by the experimental
facts such as tiny neutrino masses, unbroken $U(1)_{{\rm em}}$, etc.
However, the most general scalar potential in the model does not necessarily
lead to the Type A vacuum, because it generally has several different
local minima and sometimes the Type A minimum is not the deepest.
In Fig.\,\ref{fig:abcd} we see that among the $10^{5}$ randomly
generated samples, the Type A minimum appears as the global minimum
in about $64\%$ of them, i.e. for the remaining $36\%$ the Type
A minimum is not the deepest. If several deeper vacua coexist with
it, the Type A vacuum could be unstable since it would decay into
other deeper vacua via quantum tunneling.

Therefore, it is important to know when the Type A vacuum is the deepest
one. We have found the condition for it, given by Eqs.\,(\ref{eq:t2-31})-(\ref{eq:t2-33}).
If the potential parameters are constrained by this condition, then
the Type A minimum must be the global minimum and the stability is
guaranteed. As one can see from Fig.\,\ref{fig:togloabl}, when the
condition is partially added, the vacuum can be avoided to be of Type
B, C or D; when the condition is fully added, the vacuum must be of
Type A. An interesting physical consequence of this condition  is
that it will lead to predictions on the mass splitting of the heavy
bosons, as shown in Fig.\,\ref{fig:blue-red}.

Note that the absolute stability of the Type A vacuum  might be a
 too strong constraint on the model. Although it might decay into
a deeper minimum by quantum tunneling if it was not the deepest, the
model could still be valid since the decay rate could be low enough
so that the lifetime of the vacuum would be longer than the age of
the universe \cite{Coleman:1977py,Callan:1977pt}. That is, the vacuum
could be meta-stable. So to see whether there is the vacuum-instability
problem for a certain set of parameters, one has to compute the decay
rate. However, this could be much more complicated and is out of scope
of the current paper. For simplicity, we only consider that the Type
A minimum by itself is the deepest point of the potential so that
the decay rate will not concern us. Besides, if we obtain the absolute
stability condition, there is no longer a need to check the second
derivatives of the potential to guarantee that the extremum is a local
minimum, not a saddle point or a local maximum. Once the potential
satisfies the condition that the Type A minimum is the deepest point,
then combined with the bounded-from-below condition the Hessian matrix
must be positive (semi-)definite at the point. Therefore, though it
is a little too strong, the requirement of absolute stability may
be the simplest way to construct a valid scalar potential for the
model concerned with the above issues.
\begin{acknowledgments}
I would like to thank Sudhanwa Patra and Werner Rodejohann  for some
useful discussions and Patrick Ludl for careful reading the manuscript.
This work was partially supported by the China Postdoc Council (CPC). 
\end{acknowledgments}

\bibliographystyle{apsrev4-1}
\bibliography{ref}

\end{document}